\documentclass[10pt,twocolumn,preprintnumbers,amsmath,amssymb,nofootinbib
,superscriptaddress]{revtex4-2}
\usepackage{graphicx,longtable,mathrsfs,color,array}
\usepackage{hyperref}
\usepackage[usenames,dvipsnames]{xcolor} 
\usepackage{amssymb,amsmath,mathtools,mathrsfs,slashed} 
\usepackage{epsfig,subfigure,placeins,float} 
\usepackage{booktabs,longtable,ctable,multirow} 
\usepackage{exscale,relsize} 
\usepackage[normalem]{ulem} 
\usepackage{enumerate}
\usepackage{enumitem}
\usepackage{comment}
\usepackage{color}

\let\originalleft\left
\let\originalright\right
\renewcommand{\left}{\mathopen{}\mathclose\bgroup\originalleft}
\renewcommand{\right}{\aftergroup\egroup\originalright}

\allowdisplaybreaks[1]

\newcommand{\be}{\begin{equation}}
\newcommand{\ee}{\end{equation}}
\newcommand{\bea}{\begin{eqnarray}}
\newcommand{\eea}{\end{eqnarray}}

\newcommand{\bra}[1]{\left( #1 \right)}  
\newcommand{\brb}[1]{\left[ #1 \right]}  
  
\newcommand{\fr}[2]{\frac{#1}{#2}}

\begin{document}
\title{Black Hole Quasinormal Mode Resonances and Reconnections in Coupled Systems}

\author{Leah Jenks}
\email{ljenks3@jh.edu}
\affiliation{William H. Miller III Department of Physics \& Astronomy,\\ Johns Hopkins University, 3400 N. Charles St., Baltimore, MD 21218, USA}

\author{Hayato Motohashi}
\email{motohashi@tmu.ac.jp}
\affiliation{Department of Physics, Tokyo Metropolitan University, 1-1 Minami-Osawa, Hachioji, Tokyo 192-0397, Japan}

\author{Kazufumi Takahashi}
\email{kazufumi.takahashi@resceu.s.u-tokyo.ac.jp}
\affiliation{Research Center for the Early Universe (RESCEU), Graduate School of Science, The University of Tokyo, Tokyo 113-0033, Japan}

\author{Takuya Takahashi}
\email{takuya.takahashi@resceu.s.u-tokyo.ac.jp}
\affiliation{Research Center for the Early Universe (RESCEU), Graduate School of Science, The University of Tokyo, Tokyo 113-0033, Japan}

\author{Emanuele Berti}
\email{berti@jhu.edu}
\affiliation{William H. Miller III Department of Physics \& Astronomy,\\ Johns Hopkins University, 3400 N. Charles St., Baltimore, MD 21218, USA}

\begin{abstract}
Black hole quasinormal modes (QNMs) provide a window into the underlying structure of black holes and the environments in which they arise. QNMs have particularly rich dynamics in coupled systems beyond general relativity, in which they are known to undergo avoided crossings, resonant excitation, and mode reconnections. Recent work has shown that reconnecting QNM trajectories can trace out a distinctive double spiral structure in the complex frequency plane. In this work, we study this phenomenology in a concrete model of perturbed dynamical Chern-Simons (dCS) theory, as well as in a minimal toy model. In the former, we characterize the exceptional points, avoided crossings and reconnections for several values of the perturbation parameters, and find that the limiting frequency of the double spiral endpoints is independent of the perturbation strength. We compute the excitation factors and find that they follow the scaling characteristic of a resonance between the scalar and gravitational modes. To determine whether these phenomena depend on the specific features of dCS, we construct a toy model of coupled degrees of freedom with square potential barriers, and find that we can reproduce the avoided crossings and reconnections, indicating that these phenomena can occur more generally in coupled systems and are highly sensitive to the system parameters.
\end{abstract}

\date{\today}
\maketitle

\section{Introduction}
\label{sec:intro}

With the emergence of gravitational-wave (GW) astrophysics in the last decade, black holes (BHs) have become a laboratory with which to study fundamental physics~\cite{LIGOScientific:2016aoc, LIGOScientific:2026wfs, LIGOScientific:2026oim}. In the next decade, GW observations are expected to become a precision science, allowing for sharp tests of general relativity (GR) and other fundamental physics in the strong gravity regime~\cite{Berti:2015itd, Barack:2018yly, Yunes:2025xwp}. One particularly striking source of information about the fundamental nature of BHs comes from the ringdown of compact binary mergers. In the post-merger phase, the remnant BH is expected to emit GWs that can be represented by a combination of damped sinusoids, known as quasinormal modes (QNMs)~\cite{Kokkotas:1999bd, Nollert:1999ji, Ferrari:2007dd, Berti:2009kk, Konoplya:2011qq}. In GR, these QNMs are determined solely by the mass and spin of the BH, leading to the BH spectroscopy program~\cite{Detweiler:1980gk,Echeverria:1989hg,Finn:1992wt,Dreyer:2003bv,Berti:2005ys}. However, in the presence of beyond-GR dynamics, or other new physics, QNMs can admit a rich dynamical structure that encodes information about the underlying theory, making them a natural target to search for signatures of new physics; see, e.g., Refs.~\cite{Berti:2018vdi, Silva:2024ffz, Berti:2025hly,Chung:2025ucb,Crescimbeni:2025kxi}.

BH QNMs are represented by an infinite series of complex eigenfrequencies. The mode with the smallest damping rate is known as the fundamental mode, while those with higher damping rates are known as ``overtones''. Each QNM is characterized by its frequency and damping time, and has an associated excitation factor, which characterizes how strongly the mode is excited by a given source or other destabilizing perturbation~\cite{Leaver:1986gd,Nollert:1998ys,Berti:2006wq}. As the system parameters are varied, the QNM frequencies trace out trajectories in the complex frequency plane. When two mode trajectories closely approach each other, they can repel each other rather than crossing. This phenomenon is known as an avoided crossing and can occur near exceptional points, where both the eigenfrequencies and the corresponding eigenfunctions coalesce~\cite{Heiss:1999alz, Heiss:2012dx, Dias:2021yju, Motohashi:2024fwt,Cavalcante:2024swt,PanossoMacedo:2025xnf}. Near an exceptional point, the excitation factors can be strongly amplified, indicating resonant excitation between the modes~\cite{Motohashi:2024fwt,Yang:2025dbn}. While avoided crossings and resonant excitations can occur in gravitational QNMs for Kerr BHs, the only known avoided crossings exist for high-overtone modes, making the possibility of detection challenging~\cite{Motohashi:2024fwt,Cavalcante:2024swt, Lo:2025njp,Kubota:2026hdv}. A small perturbation to the effective potential, which may model environmental effects due to matter surrounding a BH, can lead to avoided crossings and resonances between lower overtones~\cite{Motohashi:2024fwt,Yang:2025dbn,PanossoMacedo:2025xnf}. It is also known that in coupled systems with more than one degree of freedom, the additional modes beyond the purely gravitational ones may experience this resonant behavior at lower overtone order, making them more astrophysically relevant for observations. The coupled system resonance has been identified in several settings, including in Einstein-Maxwell-axion theory, in which the gravitational, scalar, and electromagnetic modes become coupled. In this case, avoided crossings were found between gravitational and electromagnetic modes, including both the fundamental modes and overtones~\cite{Takahashi:2025uwo}.

A related but structurally distinct effect is that the modes can undergo reconnections~\cite{Motohashi:2024fwt,Cavalcante:2024swt}, or mode mixing, in which two QNM trajectories that approach each other near an exceptional point actually merge and exchange character, rather than repelling each other as in an avoided crossing. This is the BH analog of eigenvalue and eigenvector exchange that is well known in non-Hermitian systems~\cite{El-Ganainy2018,Bergholtz:2019deh,Ashida:2020dkc}. One consequence of QNM reconnection is the emergence of a double spiral structure in the QNM frequency trajectories, in which each end of the reconnected trajectory spirals toward its asymptotic frequency.
This double spiral structure has recently been found in a version of parity-violating dynamical Chern-Simons (dCS) gravity~\cite{Hu:2025efp}, raising the question of whether it is tied specifically to the parity-violating nature of dCS or can also arise in more generic coupled systems.
Related spiral trajectories have also been found in a model with a jump discontinuity connecting two P\"oschl-Teller potentials~\cite{Daghigh:2026lhk}, suggesting that such structures can arise in other settings.
These observations motivate a closer comparison of mode reconnections in dCS gravity and in a simple coupled system without assuming any specific parity-violating gravity model. 

In this work, we study QNM resonant excitation, reconnections and the double spiral structure within the context of a perturbed dCS model with a phenomenological potential bump, as well as in a simple toy model consisting of coupled degrees of freedom with square potential barriers. In a model consisting of dCS plus a phenomenological perturbation, we find an avoided crossing near an exceptional point between the fundamental scalar mode and a gravitational overtone of the full perturbed theory, closely followed in the parameter space by a reconnection that produces a mixed mode with the distinctive double spiral structure. We compute the excitation factors at the avoided crossing and verify that they are amplified, indicating a resonance that follows the characteristic ${\rm Lorentzian}^{1/4}$ scaling identified in Ref.~\cite{Motohashi:2024fwt}. We explore the dependence of the resonance and double spiral on the features of the perturbation, showing that the reconnection occurs at different points in the parameter space depending on the perturbation amplitude. We also study the evolution of the QNM spiral endpoints as $\beta$ is varied, with the dCS coupling strength scaling as $1/\beta$ in our convention. We find that, independent of the bump amplitude, as the dCS coupling parameter $\beta \rightarrow \infty$, both endpoints of the double spiral approach the fundamental scalar QNM frequency of unperturbed dCS in the $\beta\rightarrow\infty$ limit.

In order to generalize our results and determine whether the observed features rely on the specific parity-violating dCS model, we construct a minimal toy model of coupled degrees of freedom with square potential barriers, without assuming any specific parity-violating gravity model. We find that our toy model reproduces the same resonances, reconnections, and double spiral structure found in the perturbed dCS model, showing that these phenomena can occur in coupled systems beyond the specific parity-violating dCS setup. Furthermore, the double spiral is highly sensitive to the potential configuration, vanishing for very slight changes in the potentials.
 
The rest of the paper is organized as follows: in Section~\ref{sec:dCSbump}, we introduce the perturbed dCS model and define the QNM excitation factors. In Section~\ref{sec:results}, we present our results for resonant excitation, mode reconnections, and the double spiral structure, including their dependence on the perturbation amplitude and the limiting behavior of the spiral endpoints. In Section~\ref{sec:toy}, we generalize these results to a generic toy model, showing that the QNMs experience resonant avoided crossings and reconnections, and that the double spiral structure can arise without assuming any specific parity-violating gravity model. Finally, in Section~\ref{sec:conc}, we conclude with a discussion pointing to the observational relevance of these features and directions for future work. Throughout the paper, Greek indices run over the four spacetime coordinates. We work in a mostly plus metric signature and in geometrical units such that the speed of light $c = 1$, and $16\pi G = 1$, where $G$ is the gravitational constant.

\section{Perturbed dCS Model}
\label{sec:dCSbump}

In this section, we introduce the perturbed dCS model, present the coupled perturbation equations, and define the QNM excitation factors used in the analysis below. We focus on the model that has previously been studied in Ref.~\cite{Hu:2025efp}, which considers dCS gravity with the addition of a perturbing ``bump'' in the potential, which we will refer to as ``perturbed dCS''. This theory provides a backdrop to study the dynamics of a slightly more complicated coupled system than standard dCS. The additional bump perturbation in the potential can arise as a result of environmental effects from matter perturbing the system \cite{Barausse:2014tra,Barausse:2014pra,Laeuger:2025zgb} (but see Ref.~\cite{DellaRocca:2026zym} for a recent discussion on the physical viability of this interpretation), or other new beyond-GR physics. It is known that this class of bump perturbations can lead to spectral instabilities of QNMs even in GR \cite{Cheung:2021bol,Cardoso:2024mrw,Oshita:2025ibu}. We remain agnostic as to the source of the perturbation and treat the bump as a phenomenological representation of additional physics in the system.

The action of dCS gravity can be written as~\cite{Lue:1998mq, Jackiw:2003pm, Alexander:2009tp}
\begin{align}
S_{\rm dCS} = \int {\rm d}^4x \sqrt{-g} \left(R - \frac{\hat{\beta}}{2} \partial_\mu\phi\partial^\mu \phi+ \frac{\alpha}{4} \phi R\tilde{R}  \right),
\end{align} 
where $R$ is the Ricci scalar, $\phi$ is a pseudoscalar field, and $\alpha$ and $\hat{\beta}$ are coupling parameters. By convention, for $\alpha\ne 0$, one can set $\alpha=1$ by a constant rescaling of $\phi$, with the corresponding rescaling absorbed into $\hat{\beta}$. In this convention, the dCS coupling strength is determined by $1/\hat{\beta}$, and we assume $\hat{\beta}>0$ to ensure that the theory remains ghost-free. The Pontryagin density, $R\tilde{R}$, is defined as:
\be 
R\tilde{R}= \frac{1}{2}\epsilon^{\rho\sigma\alpha\beta}R^\mu{}_{\nu\alpha\beta}R^\nu{}_{\mu\rho\sigma}.
\ee 
To study the QNMs, we assume a Schwarzschild BH background, noting that the Schwarzschild BH with a trivial profile of $\phi$ is also a solution of dCS~\cite{Grumiller:2007rv}. The behavior of the QNMs in dCS is well understood~\cite{Cardoso:2009pk, Molina:2010fb,Wagle:2021tam,Chung:2025gyg,Li:2025fci}; the polar sector perturbations remain unchanged from GR, but the axial sector obtains a coupling between the gravitational perturbations, $\Psi$, and the dCS scalar field perturbations, $\Theta$. In addition to this coupling, we will consider the addition of a P\"oschl-Teller ``bump'' perturbation to the gravitational sector potential. Then, the coupled equations become
\be 
\left(\frac{{\rm d}^2}{{\rm d}r_*^2} + \omega^2\right) \begin{pmatrix}
    \Psi\\ \Theta
\end{pmatrix}
- \begin{pmatrix}
    V_{\Psi \Psi} + V_{\rm bump}  & V_{\Psi\Theta}\\
    V_{\Theta\Psi} & V_{\Theta\Theta}
\end{pmatrix}\begin{pmatrix} \Psi \\ \Theta \end{pmatrix} = 0,
\label{eq:mastereq}
\ee 
where $r_*$ is the tortoise coordinate, defined as $r_*=r+2M \log[r/(2M)-1]$, with $M$ the Schwarzschild mass parameter, and the potentials are given by:
\begin{align}
\begin{split}
    V_{\Psi\Psi}&=\bar{f}(r)\bra{\fr{\ell(\ell+1)}{r^2}-\fr{6M}{r^3}}, \\
    V_{\Theta\Theta}&=\bar{f}(r)\brb{\fr{\ell(\ell+1)}{r^2}\bra{1+\fr{36M^2}{\hat{\beta} r^6}}+\fr{2M}{r^3}}, \\
    V_{\Psi\Theta}&=\bar{f}(r)\fr{6M}{r^5}, \\
    V_{\Theta\Psi}&=\frac{(\ell+2)!}{(\ell-2)!}\frac{1}{\hat{\beta}}V_{\Psi\Theta},\\
    V_{\rm bump} &= \hat{\epsilon} \, {\rm sech}^2\left(\frac{r_*-r_a}{M}\right),
\end{split}
\end{align}
where $\hat{\epsilon}$ and $r_a$ are the amplitude and position of the bump potential, respectively, and $\bar{f}(r)$ is the usual Schwarzschild factor, $\bar{f}(r) = 1-2M/r$. Note that, in the units adopted here, $\Psi$ is dimensionless, while $\Theta$, $\hat{\beta}$, and $\hat{\epsilon}$ have dimensions~$M^2$, $M^{-4}$, and $M^{-2}$, respectively. In what follows, we use dimensionless parameters~$\beta=\hat{\beta}M^4$ and $\epsilon=\hat{\epsilon}M^2$. It should also be noted that, after an appropriate constant rescaling of $\Theta$, the potential matrix can be symmetrized such that both off-diagonal potentials scale as $\beta^{-1/2}$. Hence, in the $\beta\to\infty$ limit the system reduces to decoupled equations for the GR odd-parity gravitational mode with the bump potential and for a massless free scalar field.

Solving Eq.~\eqref{eq:mastereq} with purely ingoing boundary conditions at the horizon and purely outgoing boundary conditions at infinity yields the QNM frequencies. From these solutions, we can define the excitation factors, which describe how a QNM responds to an external source or perturbation. For a coupled system with $N$ degrees of freedom, the excitation factor~$\bar{B}_{n}$ for a mode~$n$ can be defined as~\cite{Takahashi:2025uwo}
\be 
\bar{B}_{n} = \frac{i {\rm det} \mathbf{A}^{{\rm out}}}{(2i\omega_{n})^N}\left(\frac{{\rm d}}{{\rm d}\omega} {\rm det}\mathbf{A}^{\rm in}\right)^{-1}\Bigg\rvert_{\omega = \omega_{n}}.
\ee 
Here, $\mathbf{A}^{\rm out}$ and $\mathbf{A}^{\rm in}$ are the $N\times N$ matrices whose columns contain, respectively, the outgoing and ingoing amplitudes at infinity of $N$ linearly independent solutions normalized to purely ingoing waves at the horizon. Their determinants can be computed from the corresponding Wronskian determinants as described in Ref.~\cite{Takahashi:2025uwo}. Note that the excitation factor~$\bar B_n$ has dimensions~$M^{N-1}$. In what follows, we use the dimensionless excitation factor defined by $B_n=\bar B_n/M^{N-1}$. In the present system, $N=2$.

\section{Resonant Excitation, Reconnections, and Double Spiral Structure}
\label{sec:results}

In this section, we present numerical results for resonant excitation, mode reconnections, and the double spiral structure in the perturbed dCS model introduced in Section~\ref{sec:dCSbump}. We examine their dependence on the bump amplitude and the evolution of the spiral endpoints as the dCS coupling becomes weak.

In Ref.~\cite{Hu:2025efp}, the authors noted that the QNM migration trajectories for the scalar and gravitational modes in this theory undergo mode reconnections, depending on the coupling, $\beta$. Here, we explore this behavior in further detail. 
We determine $\omega$ by matching the mode functions integrated from the boundary conditions constructed using asymptotic expansions at infinity and at the horizon. For details, see, e.g., Ref.~\cite{Takahashi:2025uwo}. In Ref.~\cite{Hu:2025efp}, the authors showed that there is a reconnection between scalar and gravitational mode trajectories for some value of the coupling between $\beta = 4$ and $\beta=5$. In particular, the reconnection leads to a distinctive double spiral structure in the QNM trajectories. In order to more precisely characterize the behavior of the QNM trajectories, and to understand the exceptional points, resonances, and reconnections, we focus on this region of the parameter space.

\begin{figure}[t]
     \includegraphics[width=0.48\textwidth]{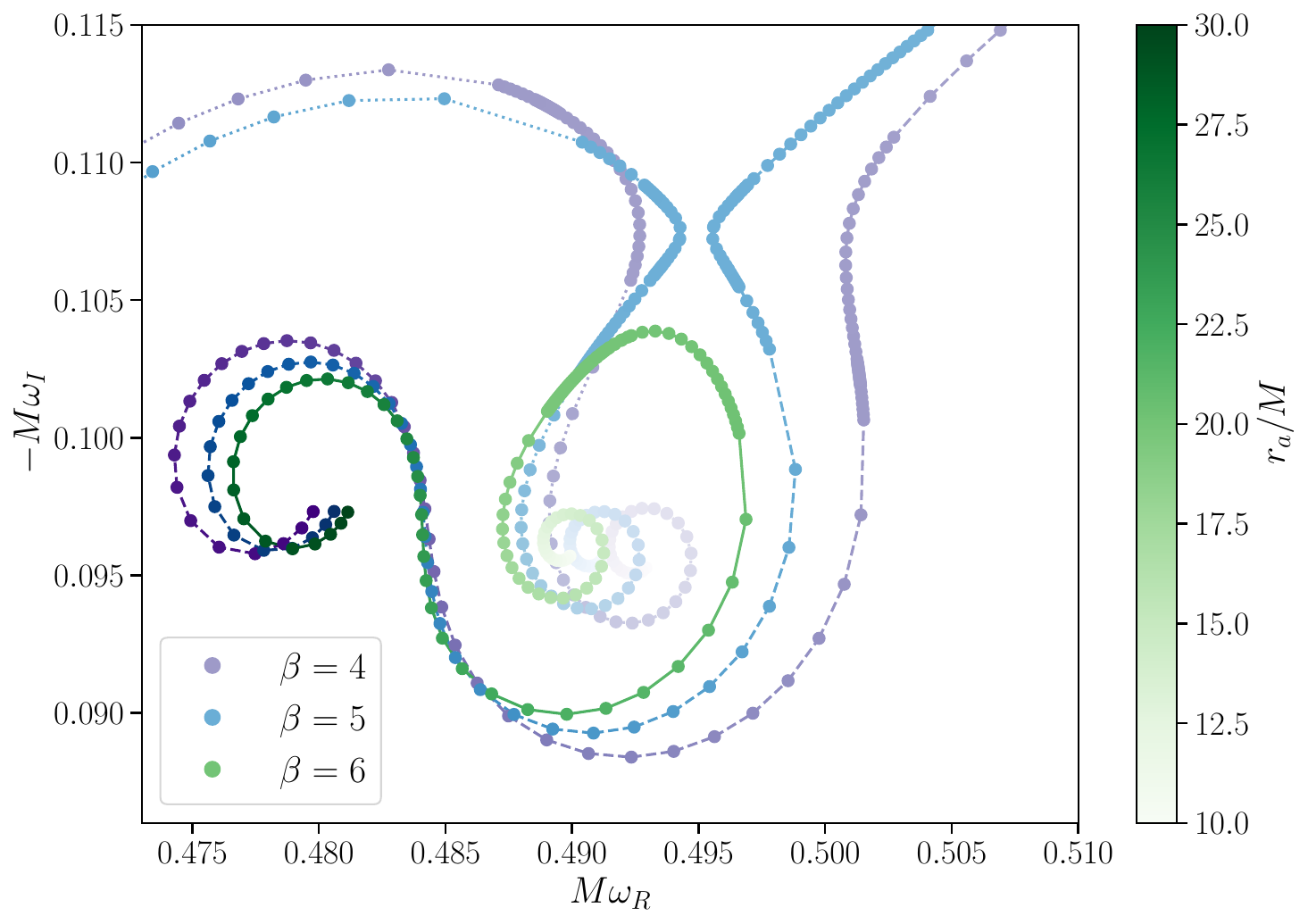}
    \caption{Scalar (dotted) and gravitational (dashed) QNM trajectories approaching each other (purple, $\beta = 4$), experiencing an avoided crossing (blue, $\beta = 5$) and undergoing a reconnection (green, $\beta= 6$) near an exceptional point at $r_a \approx 19.9\,M$. The color gradient from light to dark corresponds to low to high $r_a$. The amplitude of the bump is set by taking $\epsilon=10^{-2}$.}
    \label{fig:eps1em2avoided}
\end{figure}

We begin by resolving the mode trajectories near $\beta=5$, where our calculations locate the reconnection between a scalar mode and a gravitational mode that is an overtone of the full perturbed dCS theory. Figure~\ref{fig:eps1em2avoided} shows the mode trajectories near the reconnection, as a function of bump distance, $r_a$, for $\beta = 4$ (purple), $\beta = 5$ (blue), and $\beta = 6$ (green). We write the QNM frequency as $\omega=\omega_R+i\omega_I$, where $\omega_R$ and $\omega_I$ denote its real and imaginary parts, respectively. The color bar is representative of all three values of $\beta$, where a lighter shade denotes smaller $r_a$. Taking $\ell = 2$ and $\epsilon = 10^{-2}$, we see that the scalar (dotted) and gravitational (dashed) modes begin to approach each other at $\beta = 4$ and experience an avoided crossing at $\beta \simeq 5$, at $r_a \approx 19.9\,M$, near an exceptional point. Then, beyond $\beta= 5$, the two mode trajectories reconnect, giving rise to the characteristic double spiral structure. The reconnection joins the low-$r_a$ end of one original trajectory to the high-$r_a$ end of the other, forming a double spiral. The double spiral mode is not purely scalar or gravitational, but rather a mixture of the two.

\begin{figure}[t]

     \includegraphics[width=0.48\textwidth]{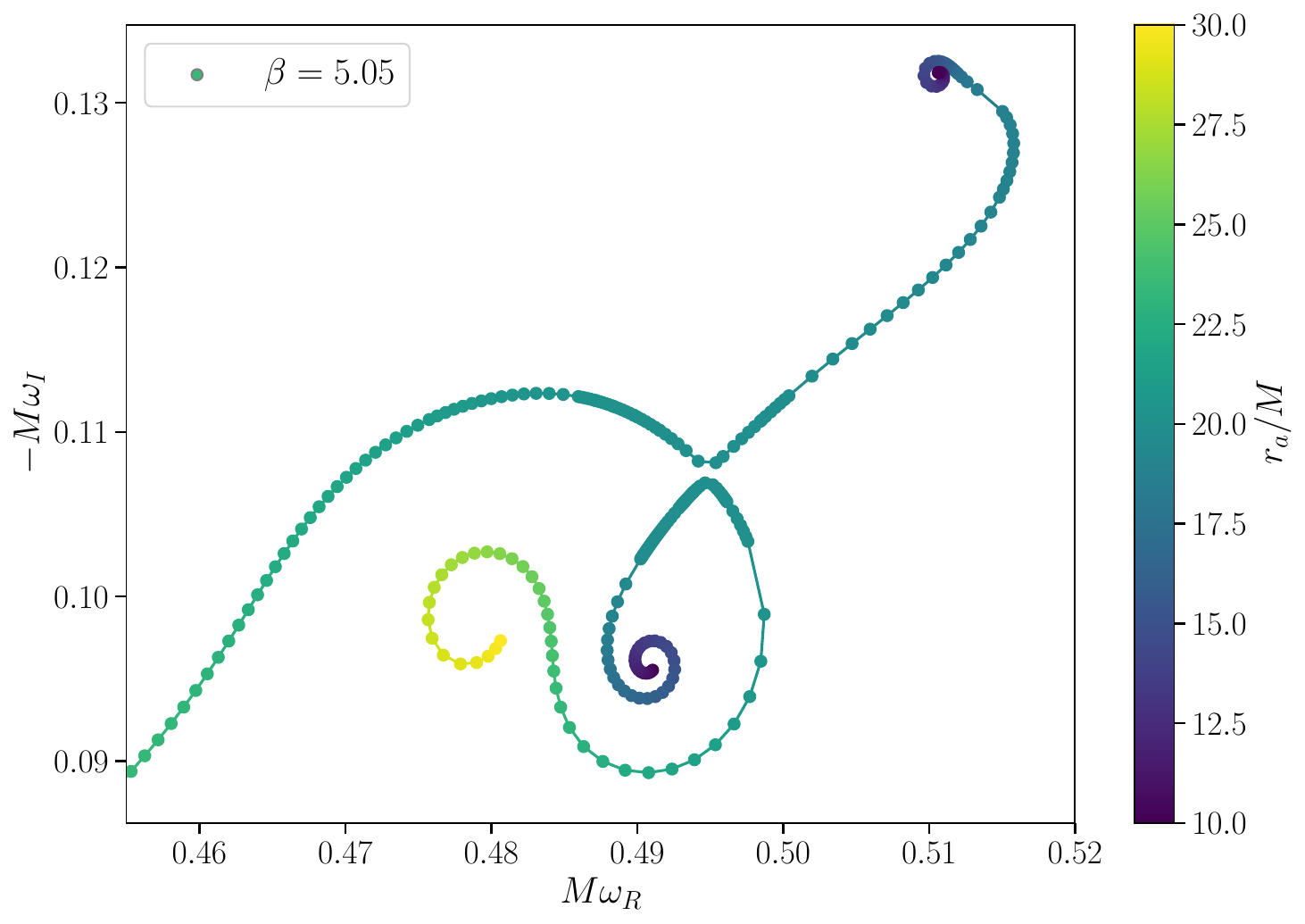}
    \caption{Mode reconnection of the scalar and gravitational modes at $\beta = 5.05$, leading to two new mode branches. }
    \label{fig:eps1em2-recon}
\end{figure}

The same reconnection joins the remaining ends of the two original trajectories to form a second mixed-mode trajectory. To resolve the reconnection within $5<\beta<6$, Fig.~\ref{fig:eps1em2-recon} shows the two mixed-mode trajectories at $\beta=5.05$, by which point the reconnection has already occurred. One trajectory goes to the origin at large $r_a$, and spirals to a point at small $r_a$. The second mixed mode has the distinctive double spiral structure seen in Fig.~\ref{fig:eps1em2avoided}. The two new mixed modes continue to experience an avoided crossing near $r_a \approx 19.9\, M$.

\begin{figure}[htb!]
    \includegraphics[width=0.48\textwidth]{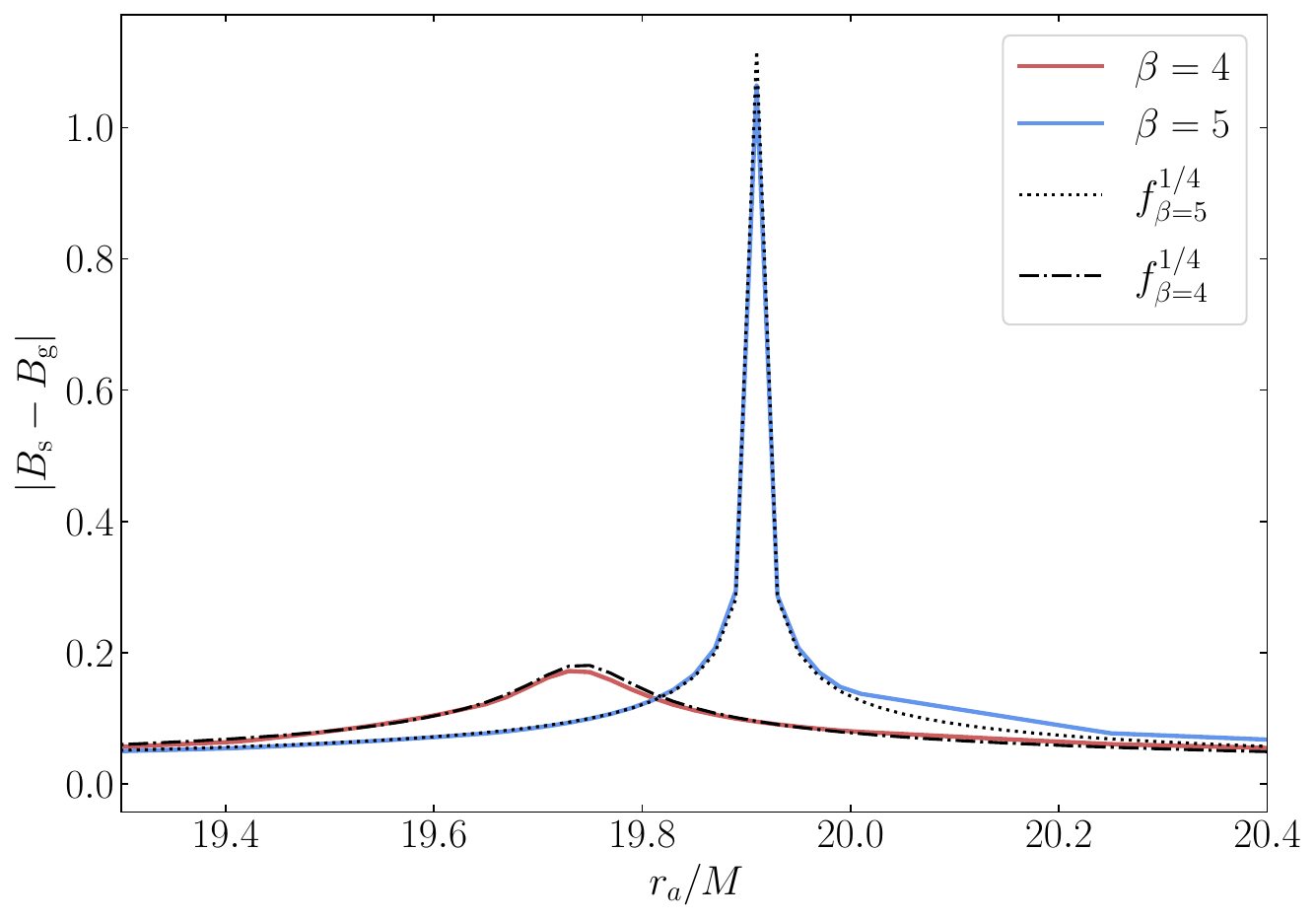}
    \includegraphics[width=0.48\textwidth]{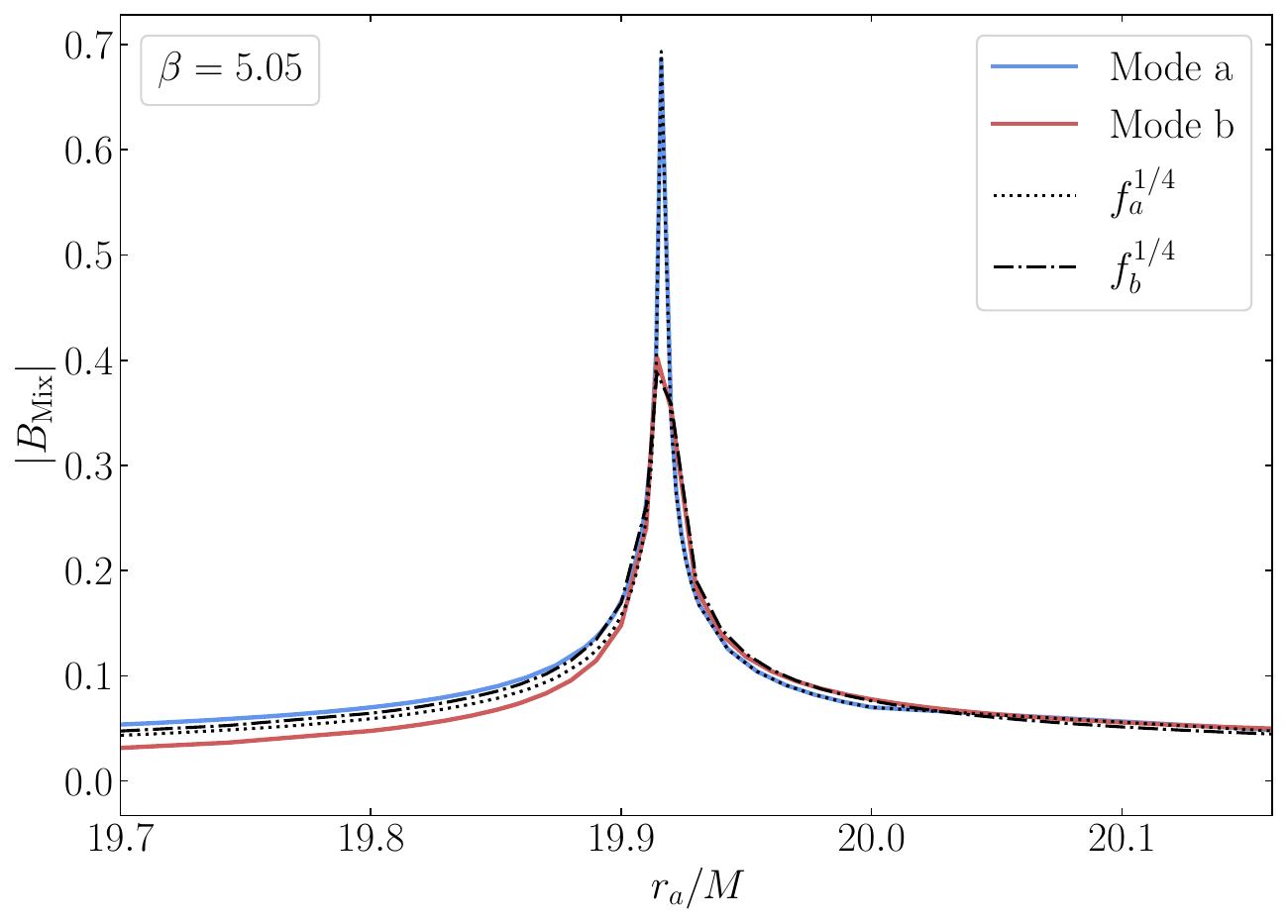}
    \caption{Excitation factors at the avoided crossing near the exceptional point for the scalar and gravitational modes before reconnection (top) and the mixed modes after reconnection (bottom). In both cases, we see the excitation factors indicate a resonance. In both cases we show the best fit to the quarter power of a Lorentzian function.}
    \label{fig:eps1em2-excitation}
\end{figure}

The avoided crossing near $r_a \approx 19.9\,M$ is present before and after the reconnection, suggesting proximity to an exceptional point. We characterize the associated resonance by computing the excitation factors near the avoided crossing, shown in Fig.~\ref{fig:eps1em2-excitation}. In the top panel, we compute the magnitude of the difference between the excitation factors for the scalar and gravitational modes, comparing $\beta = 4$ and $\beta = 5$. Following our intuition from Fig.~\ref{fig:eps1em2avoided}, the peak of $|B_{\mathrm{s}} - B_{\mathrm{g}}|$ for $\beta = 5$ is nearly a factor of five larger than that for $\beta = 4$, indicating a resonance near the exceptional point that is amplified as the modes reach their closest approach in the avoided crossing. In the bottom panel, we show $|B_{\rm mix}|$ for both mixed modes after reconnection for $\beta = 5.05$, where we refer to the double spiral mode as ``Mode a'' and the secondary reconnection as ``Mode b''. Again, the peak of the excitation factors indicates that even after reconnection, there is still resonant behavior between the modes. In both panels, we additionally show the best analytic fit to $|B_n|$. It was previously shown in Ref.~\cite{Motohashi:2024fwt} that the excitation factors scale as a power of a Lorentzian function,
\be 
f(r_a) = \frac{f_0}{(r_a - r_0)^2 + \gamma^2}.
\label{eq:fL}
\ee 
Here, $f_0$ sets the normalization, $r_0$ is the peak position, and $\gamma$ is the half-width at half-maximum of $f$, with $f(r_0)=f_0/\gamma^2$. Note that $f_0$, $r_0$, and $\gamma$ have dimensions~$M^2$, $M$, and $M$, respectively. For Kerr BHs in GR, the excitation factors scale as $f^{1/4}$~\cite{Motohashi:2024fwt}, while for BHs in coupled Einstein-Maxwell-axion theory, the QNM excitation factors scale as $f^{1/2}$~\cite{Takahashi:2025uwo}. In the present case, we find that $|B_{\rm mix}|$ and $|B_{\mathrm{s}} - B_{\mathrm{g}}|$ both fit well to $f^{1/4}$, shown in black, aligning well with the known Kerr result. We give the specific fitting parameters in Table~\ref{tab:B_fit_Fig3} of Appendix~\ref{app:dCS-Excitation}.

The difference from the $f^{1/2}$ scaling found in the Einstein-Maxwell-axion system reflects the different QNM frequency trajectories, rather than the number of coupled fields. As shown in Ref.~\cite{Motohashi:2024fwt}, the resonant excitation factors approximately satisfy $B_{\mathrm{s}}-B_{\mathrm{g}}\propto(\omega_{\mathrm{s}}-\omega_{\mathrm{g}})^{-1}$. In the present case, the frequency trajectories near the avoided crossing approximately follow the hyperbola, as in the Kerr case. Through this inverse relation, the corresponding excitation-factor trajectories trace lemniscates of Bernoulli, giving rise to the $f^{1/4}$ profile~\cite{Motohashi:2024fwt}. The different frequency evolution in the Einstein-Maxwell-axion system instead leads to the $f^{1/2}$ profile~\cite{Takahashi:2025uwo}.

\begin{figure*}[htb!]
\includegraphics[width=\textwidth]{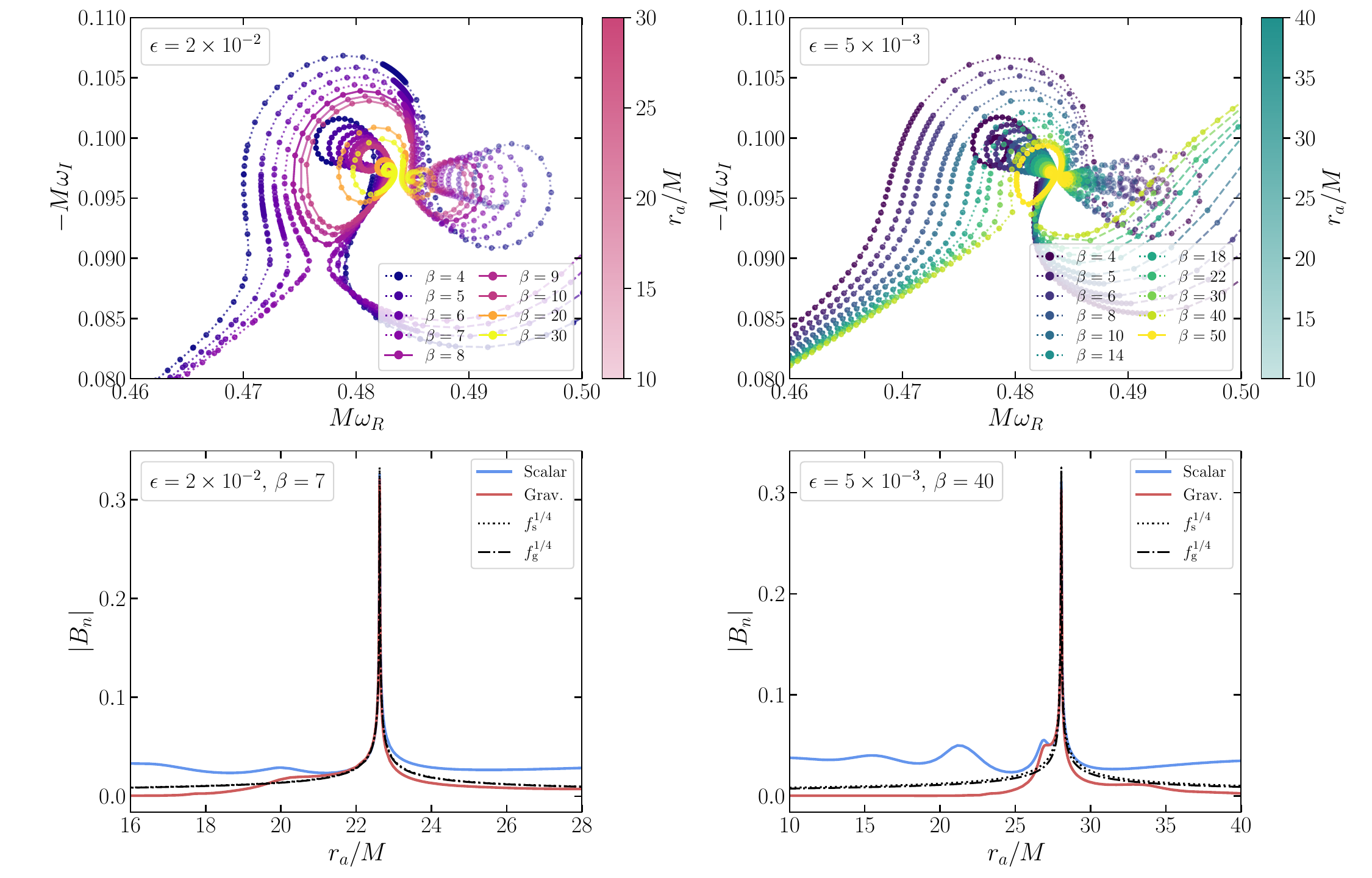}
\caption{QNM trajectories for varying $\epsilon$. For illustration, we consider $\epsilon = 2\times 10^{-2}$ (top left) and $\epsilon=5\times 10^{-3}$ (top right). We show the scalar (dotted), gravitational (dashed) and mixed (solid) mode double spiral. We then show the excitation factors for both values of $\epsilon$ at their closest avoided crossings; $\beta = 7$ for $\epsilon = 2\times 10^{-2}$ and $\beta = 40$ for $\epsilon=5\times 10^{-3}$ (bottom left and right, respectively), along with the best fit to the quarter power of a Lorentzian function for each curve. }
\label{fig:varyamplitude}
\end{figure*}

In the above discussion, we have focused on the effects of varying $\beta$ and $r_a$; however, the amplitude of $V_{\rm bump}$, $\epsilon$, also controls the dynamics of the QNMs. Therefore, we now explore how changing the amplitude of $V_{\rm bump}$ impacts the resonances and reconnections of the QNM trajectories. We consider a factor of two increase and decrease of the bump amplitude to $\epsilon = 2 \times 10^{-2}$ and $\epsilon = 5 \times 10^{-3}$ and focus on the same two modes as in Fig.~\ref{fig:eps1em2avoided} to compare directly. The results for both of these cases are shown in Fig.~\ref{fig:varyamplitude}, where we show the QNM mode trajectories and excitation factors at the avoided crossings in the upper and lower panels, respectively. In the upper panels, we again choose a representative value for the color bar, noting that lighter shades correspond to smaller $r_a$.

First, consider the high-$\epsilon$ case. The top left panel shows the same two scalar and gravitational modes as in Fig.~\ref{fig:eps1em2avoided} for increasing $\beta$. There is again an avoided crossing that emerges as $\beta$ increases, but now at a different point in both $\beta$ and $r_a$ than was observed previously for $\epsilon = 10^{-2}$ in Fig.~\ref{fig:eps1em2avoided}. The avoided crossing now occurs at $r_a \approx 22.6\, M$ and the two modes are at their closest approach for $\beta \simeq 7$, after which a reconnection occurs prior to $\beta = 8$, leading to the double spiral. Beyond $\beta = 8$, the double spiral structure remains, but continues to grow smaller. To characterize the resonance, we again compute the excitation factors for each mode at $\beta = 7$ and find a sharp peak near the exceptional point at $r_a \approx 22.6\, M$. The excitation factors can once again be fit using Eq.~\eqref{eq:fL}, with an $f^{1/4}$ scaling at the peak. The best-fit values are shown in Table~\ref{tab:B_fit_Fig4} of Appendix~\ref{app:dCS-Excitation}.

We observe similar behavior in the low-$\epsilon$ case, shown on the right. In the top panel, we see the two modes drift towards each other much more gradually in $\beta$, where the closest avoided crossing occurs at $\beta = 40$ and the reconnection into the double spiral occurs by $\beta = 50$. The avoided crossing again occurs at a larger value of $r_a$, namely $r_a \approx 28 \,M$. The peaks of the excitation factors, shown in the bottom panel, again fit to $f^{1/4}$, as shown in Table~\ref{tab:B_fit_Fig4} of Appendix~\ref{app:dCS-Excitation}. Notice that in both cases, mode~a continues to have an enhanced excitation factor beyond the avoided crossing. This is likely due to another untracked mode being nearby in the frequency space, leading to low-level excitation.

Another notable result that can be appreciated from Fig.~\ref{fig:varyamplitude} is that the values of $r_a$ and $\beta$ near reconnection vary nonmonotonically with $\epsilon$ over the cases studied. Both raising and lowering the perturbation amplitude, $\epsilon$, from the benchmark~$\epsilon = 10^{-2}$ case increases the critical value of $r_a$ where the avoided crossing occurs, and also increases the $\beta$ value where the modes undergo the reconnection. Reference~\cite{Hu:2025efp} reported that the critical bump position for mode overtaking increases as the amplitude decreases. This threshold concerns a change in the dominant mode and is distinct from the avoided-crossing and reconnection locations considered here. We defer a more detailed study of their dependence on $\epsilon$ to future work.

\begin{figure}[t]
        \includegraphics[width=0.48\textwidth]{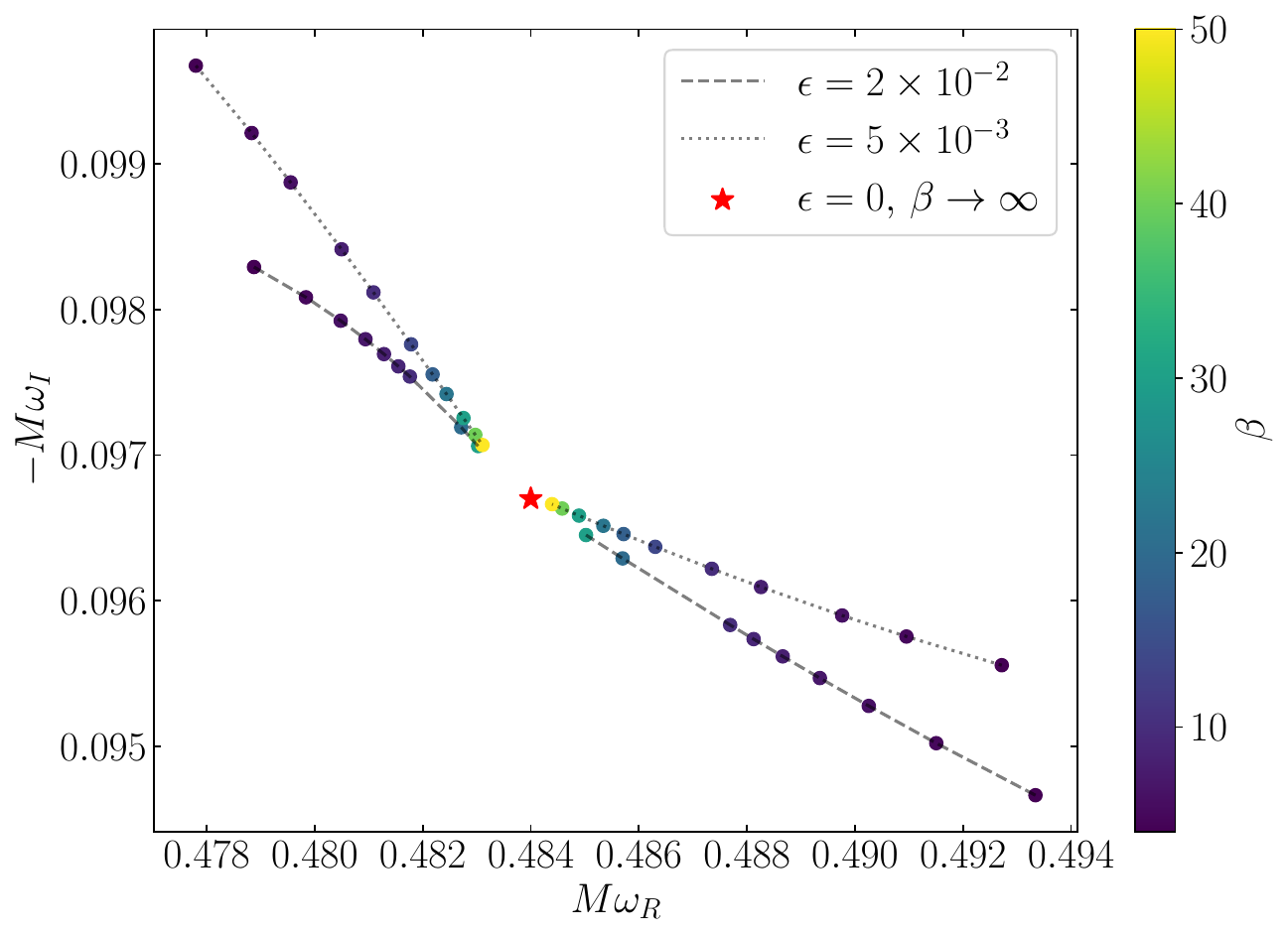}
        \caption{Evolution of the spiral end points as $\beta \rightarrow \infty$ for both the $\epsilon=2\times 10^{-2}$ (dashed) and $\epsilon = 5\times 10^{-3}$ (dotted) trajectories. At large $\beta$ the initial points approach the $\beta \rightarrow \infty$ limit of the unperturbed dCS ($\epsilon=0$) scalar frequency, denoted by the red star.}
        \label{fig:initialpts}
\end{figure}

To understand the final fate of these QNM trajectories, let us further characterize the mode evolution. From the trajectories in the upper row of Fig.~\ref{fig:varyamplitude}, we can appreciate that as $\beta$ increases after the reconnection occurs, the spiral begins to shrink. To see this precisely, consider the endpoint evolution of each end of the spiral as a function of $\beta$. In Fig.~\ref{fig:initialpts}, we show the trajectories of each spiral endpoint for $\epsilon = 2 \times 10^{-2}$ (dashed) and $\epsilon = 5 \times 10^{-3}$ (dotted) as a function of increasing $\beta$, corresponding to the values in the upper row of Fig.~\ref{fig:varyamplitude}. We find that for both $\epsilon$ cases, the endpoints approach the frequency of the fundamental scalar mode of the $\beta\rightarrow \infty$ limit of the standard dCS theory, given by~\cite{Molina:2010fb}
\be 
M \omega_\infty = 0.484 - 0.0967i,
\label{eq:omegainf}
\ee 
which is marked with a red star in the figure. These results indicate that, for the configurations studied, all spiral endpoints converge to the same \textit{unperturbed} dCS scalar QNM frequency as $\beta$ increases, independently of the bump amplitude and their starting positions in frequency space, rather than approaching a bump-dependent value.

\section{Toy Model}
\label{sec:toy}

To generalize the features observed in the previous section to coupled systems beyond the perturbed dCS theory, let us study a simple toy model. We explore whether the double spiral structure can arise in a generic coupled system without assuming any specific parity-violating gravity model, and how it responds to changes in the potential configuration.

To this end, let us consider a minimal model of a coupled system with matrix-valued effective potential
\be 
V_{\rm toy} = \begin{pmatrix} V_{11} + V_{\rm pert} & V_{12} \\ V_{21} & V_{22}\end{pmatrix},
\ee 
where each of the potential components is given by a top-hat function,
\be 
V_{i}(x) = h_i H(x - x_i^{\rm L})H(x^{\rm R}_i - x),
\ee 
for $i \in \{11, 22, 12, {\rm pert}\}$, with $V_{21}=V_{12}$, where $H$ is the Heaviside step function. The right- and left-hand edges of each square barrier are characterized by $x^{\rm R}$ and $x^{\rm L}$, respectively, with amplitude~$h_i$. In what follows, all quantities in the toy model are dimensionless, and we denote the left edge of $V_{\rm pert}$ by $x_a$, i.e., $x_a=x^{\rm L}_{\rm pert}$.

We begin with a base scenario, taking the height and width of each barrier from the peak value and full width at half maximum of the perturbed dCS potential configuration at $\beta =4$, and assume that $V_{\rm pert}$ has a width of one. In doing so, the dCS potential and tortoise coordinate are made dimensionless by expressing them in units of $M^{-2}$ and $M$, respectively. We refer to these as $V_i^{\rm base}$, and the particular numbers for $h_i, x^{\rm R}_i$, and $x_i^{\rm L}$ can be found in Appendix~\ref{app:Toy}. The configuration with $V_i^{\rm base}$ reproduces the double spiral. We then shift the relative position of the potential barriers to determine whether the reconnection remains. There are a wide range of potential configurations such that the mode mixing remains, but we find that it is highly sensitive to the barrier positions, heights, and widths.

\begin{figure*}[htb!]
 \includegraphics[width=\textwidth]{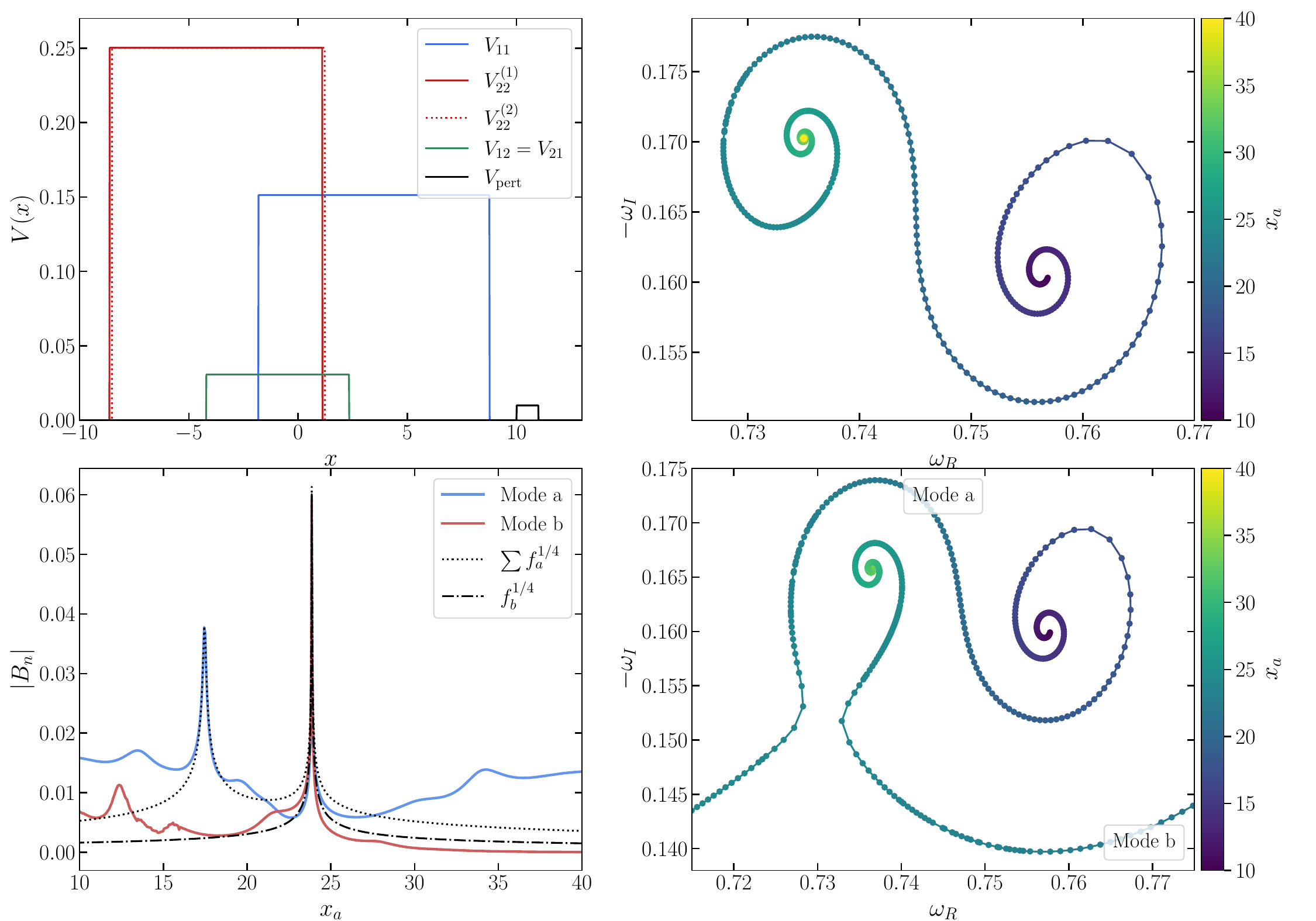}

      \caption{Toy model potential configurations (top left), QNM trajectories for configuration~$1$ (top right) and configuration~$2$ (bottom right), and excitation factors near the avoided crossing in configuration~$2$ (bottom left).}
      \label{fig:QNMs-Toy}
\end{figure*}

As a concrete example, consider a shift to the left in the position of the barrier of $V_{22}$ from the base case, as shown in the top left panel of Fig.~\ref{fig:QNMs-Toy}. We show two examples, which we label as ``$1$'' and ``$2$''. In configurations~$1$ (solid red) and $2$ (dotted red), the $V_{22}$ barrier is shifted to the left by $5.90$ and $5.79$, respectively; we denote the shifted potentials by $V_{22}^{(1)}$ and $V_{22}^{(2)}$. We also show the base values for $V_{11}$ in blue, the coupling potential, $V_{12} = V_{21}$ in green, and an example location for $V_{\rm pert}$ in black.

This incremental change between the two configurations has a drastic impact on the two modes. For configuration~$1$, as $x_a$ is varied, we observe that a double spiral emerges, as shown in the top right panel of Fig.~\ref{fig:QNMs-Toy}, exhibiting similar behavior to that in the perturbed dCS case. Again the double spiral appears as a reconnection of two individual modes with endpoint spirals. However, with the small shift to configuration~$2$, the double spiral vanishes and we recover the two individual modes at their avoided crossing point, shown in the bottom right panel of Fig.~\ref{fig:QNMs-Toy}, which we label ``mode~a'' and ``mode~b''. In this case, the avoided crossing occurs near an exceptional point at $x_a \approx 24$. 

We further compute the excitation factors near the avoided crossing and find that $|B_{\mathrm{a}}|$ and $|B_{\mathrm{b}}|$ peak at approximately the same value of $x_a$, as shown in the bottom left panel. For the toy model, we use the Lorentzian function~$f(x_a)=f_0/[(x_a-x_0)^2+\gamma^2]$, where $f_0$, $x_0$, and $\gamma$ are dimensionless constants. For mode~b, the excitation factor peak at $x_a \approx 24$ is again well fit by the characteristic quarter power Lorentzian, $f^{1/4}$, scaling~\cite{Motohashi:2024fwt}, with the specific fits given in Appendix~\ref{app:Toy}. Similar to the excitation factors in the perturbed dCS case, the tails do not go cleanly to zero. This is even more pronounced in mode~a, which has a clear second peak near $x_a = 17.5$, which also has an $f^{1/4}$ scaling. We thus fit the magnitude of the excitation factor for mode~a with a sum of two quarter-power Lorentzian profiles. This double-peaked structure suggests that mode~a undergoes a resonance with mode~b near $x_a\approx24$, as well as with another nearby mode that we do not track near $x_a\approx17.5$.

From our exploration of the toy model, we can appreciate the sensitivity of the double spiral structure, given that the modes undergo reconnections for percent level changes in the potential configuration. An additional takeaway from our study of the coupled square barrier model is that the emergence of the double spiral does not appear to rely on the specific parity-violating structure of the dCS setup. Our toy model is not based on any specific parity-violating gravity theory, yet we continue to observe the double spiral structure of the QNMs. This shows that the reconnections and double spiral structure observed here can arise more generally in coupled systems.

\section{Discussion and Conclusions}
\label{sec:conc}

In this paper we have discussed the exceptional points, avoided crossings, resonances, and reconnections of BH QNM trajectories in coupled systems. We have focused on two case studies: perturbed dynamical Chern-Simons gravity and a toy model of two coupled square barriers with a perturbation. In both models, we have found avoided crossings and reconnections into a double spiral structure, and have computed the excitation factors to characterize the associated resonant excitation near exceptional points. In the perturbed dCS case, the values of $r_a$ and $\beta$ near reconnection vary nonmonotonically with $\epsilon$ over the cases studied. We further showed that, for the perturbation amplitudes studied, both spiral endpoints approach the same fundamental scalar QNM frequency of the unperturbed dCS theory as $\beta\to\infty$. The toy model also exhibits an avoided crossing and a reconnection into a double spiral. In this case, incremental changes in the coupled barrier configuration drastically change the behavior of the QNMs. Furthermore, this confirms that these phenomena are not specific to the parity-violating dCS setup.

There are many avenues for future exploration. It would be enlightening to understand more concretely the full range of systems that these phenomena can occur in, and the conditions for the reconnection and appearance of a double spiral rather than just an avoided crossing between two modes. Furthermore, it would be interesting to study these phenomena within the context of observations and determine if the double spiral reconnection leads to unique signatures in ringdown waveforms.

\acknowledgements
L.J.\ is supported by the Provost's Postdoctoral Fellowship at Johns Hopkins University. 
H.M.\ is supported in part by JSPS (Japan Society for the Promotion of Science) KAKENHI Grant No.~JP22K03639 and No.~JP26K21813.
K.T.\ is supported in part by JSPS KAKENHI Grant No.~JP23K13101.
T.T.\ is supported by JSPS KAKENHI Grant No.~JP25KJ0067 and No.~JP25K17397.
E.B.\ is supported by NSF Grants No.~AST-2606672, No.~PHY-2513337, No.~PHY-090003, and No.~PHY-20043, by the Simons Foundation [MPS-SIP-00001698, E.B.], by the Simons Foundation International [SFI-MPS-BH-00012593-02], and by Italian Ministry of Foreign Affairs and International Cooperation Grant No.~PGR01167.

\appendix
\section{Perturbed dCS Excitation Factor Fitting Parameters}
\label{app:dCS-Excitation}
In this appendix we list the parameters of the Lorentzian function $f$ used in the $f^{1/4}$ fits to the excitation factors in the perturbed dCS model shown in Figs.~\ref{fig:eps1em2-excitation} and \ref{fig:varyamplitude}.
The best fits to the excitation factors in Fig.~\ref{fig:eps1em2-excitation} are shown in Table~\ref{tab:B_fit_Fig3}, while
those in Fig.~\ref{fig:varyamplitude} are listed in Table~\ref{tab:B_fit_Fig4}.\\

\begin{table}[htb]
\centering
\caption{Best-fit parameters for $|B_{\mathrm{s}} - B_{\mathrm{g}}|$ and $|B_{\rm Mix}|$ near the avoided crossing shown in Fig.~\ref{fig:eps1em2-excitation}. }
\label{tab:B_fit_Fig3}
\begin{tabular}{lccc}
\toprule
 & $f_0/M^2$ & $\gamma/M$ & $r_0/M$ \\
\midrule
$|B_{\mathrm{s}} - B_{\mathrm{g}}|_{\beta = 4}$      & $2.56\times10^{-6}$ & 0.0483 & 19.74 \\
$|B_{\mathrm{s}} - B_{\mathrm{g}}|_{\beta = 5}$      & $2.56\times10^{-6}$ & 0.00128 & 19.91 \\
$|B_{\rm mix}|_{\text{mode~a}}$ & $1.66\times10^{-7}$ & $6.70\times10^{-4}$ & 19.91 \\
$|B_{\rm mix}|_{\text{mode~b}}$ & $2.34 \times 10^{-7}$ & 0.0020 & $19.92$ \\
\bottomrule
\end{tabular}
\end{table}

\begin{table}[htb]
\centering
\caption{Best-fit parameters for $|B_n|$ near the avoided crossings shown in the bottom row of Fig.~\ref{fig:varyamplitude}.}
\label{tab:B_fit_Fig4}
\begin{tabular}{lccc}
\toprule
 & $f_0/M^2$ & $\gamma/M$ & $r_0/M$ \\
\midrule
$\beta = 7$, scalar      & $2.56\times10^{-7}$ & 0.00346 & 22.63 \\
$\beta = 7$, grav.      & $2.34\times10^{-7}$ & 0.00417 & 22.63 \\
$\beta = 40$, scalar    & $1.38\times10^{-6}$ & $0.0050$ & 28.05 \\
$\beta = 40$, grav.     & $8.88 \times 10^{-7}$ & $0.0047$ & 28.05 \\
\bottomrule
\end{tabular}
\end{table}

\section{Toy Model Parameters}
\label{app:Toy}
In this appendix we provide the parameters for the toy model base potential configuration, as well as the best-fitting parameters for the excitation factors. 
The parameters of the base toy model are listed in Table~\ref{tab:toy_params}.
The best-fit parameters for the excitation factors in Fig.~\ref{fig:QNMs-Toy} are given in Table~\ref{tab:B_fit_Toy}.

\begin{table}[htb]
\centering
\caption{Base parameters for the toy model potential components.}
\label{tab:toy_params}
\begin{tabular}{lccc}
\toprule
 & $h$ & $x^{\rm R}$ & $x^{\rm L}$ \\
\midrule
$V_{11}$       & 0.151287  & 8.77135     & $-1.80851$ \\
$V_{22}$       & 0.250242  & 7.01288     & $-2.72722$ \\
$V_{12}$       & 0.0307623 & 2.34334     & $-4.19977$ \\
$V_{\rm pert}$ & $10^{-2}$ & $x_a + 1$ & $x_a$ \\
\bottomrule
\end{tabular}
\end{table}

\begin{table}[htb]
\centering
\caption{Best-fit parameters for $|B_n|$ in the toy model, shown in the bottom left panel of Fig.~\ref{fig:QNMs-Toy}.}
\label{tab:B_fit_Toy}
\begin{tabular}{lccc}
\toprule
 & $f_0$ & $\gamma$ & $x_0$ \\
\midrule
Mode a, peak 1     & $1.08\times10^{-8}$ & 0.0800 & 17.46 \\
Mode a, peak 2     & $1.07\times10^{-9}$ & 0.00802 & 23.87 \\
Mode b     & $1.30\times10^{-9}$ & 0.00802 & 23.87 \\
\bottomrule
\end{tabular}
\end{table}

\bibliography{bib}

\end{document}